\documentstyle[aps,prl,amssymb,graphics]{revtex}

\begin{document}

\draft
\widetext
\narrowtext
\twocolumn
\wideabs{
\title{Enhanced Estimation of a Noisy Quantum Channel Using Entanglement}

\author{Dietmar G. Fischer, Holger Mack, Markus A. Cirone and
Matthias Freyberger} 
\address{Abteilung f\"ur Quantenphysik, Universit\"at Ulm, 
D-89069 Ulm, Germany} 


\maketitle
\begin{abstract}
We discuss the estimation of channel parameters for a noisy quantum channel --
the so-called Pauli channel -- using finite resources. It turns out that prior
entanglement considerably enhances the fidelity of the estimation when we
compare it to an estimation scheme based on separable quantum states.
\end{abstract}
}

\section{Introduction}
In the last few years, the field of quantum information processing has made
enormous progress. It has been shown that the laws of quantum mechanics open
completely new ways of communication and computation \cite{zeilinger}. This
progress has mainly been driven by understanding more and more about the physics
of entanglement. The corresponding nonclassical correlations are central to
completely new applications like quantum cryptography using entangled systems
\cite{crypto},
teleportation \cite{teleport} and dense coding \cite{denscode}. These are all
examples of superior information transmission with quantum mechanical means. 
\par
Consequently, much work has been done toward the understanding of quantum
communication channels. In principle a quantum channel is
simply a transmission line between a sender, say Alice, and a receiver, say Bob,
that allows them to transfer quantum systems. The noiseless channel leaves the
quantum states of the transmitted systems intact. In other words, such a channel
is completely isolated from any environment. This is certainly a strong
idealization. More realistic is the noisy quantum channel that takes into
account the interaction of the sent system with an environment: the
corresponding quantum state decoheres. In effect this process can be described
by a superoperator $C$ which in general maps Alice's pure state
$|\psi\rangle\langle\psi|$
on a density operator $\hat{\rho}=C(|\psi\rangle\langle\psi|)$ on Bob's
side \cite{kraus}.\par
The prominent topics of research in the field of quantum channel theory are to
understand the notion of capacity of a quantum channel and to understand the
role played by entanglement. The proof \cite{schumacher} of the quantum analog
of Shannon's noiseless coding theorem \cite{shannon} 
was a milestone in this field indicating
that a quantum theory of information transmission is possible in parallel to its
classical counterpart. Consequently, much work then concentrated on the concept
of capacity for a general noisy quantum channel \cite{schuma1,capacity}. 
Hence these investigations aim
at quantifying the maximum rate at which information can be sent through a noisy
quantum channel: in analogy to the noisy-channel coding theorem of Shannon
\cite{shannon}. Moreover, it was shown \cite{schuma1} that entanglement can be
used as a resource to quantify the noise of a quantum channel. \par
However, since a quantum channel can carry classical as well as quantum
information, several different capacities can be defined 
\cite{schuma1,capacity}. 
In particular, it is
well-known that prior entanglement can enhance the capacity of quantum channels
to transmit classical information \cite{denscode,transclass}. 
It is therefore justified to consider
entanglement as a fruitful resource which has no classical analogue
\cite{analogue}.\par
In the present paper we shall discuss a different application of
entanglement in the field of quantum channels. Let us suppose that Alice and Bob
are connected by a specific noisy channel. Both parties know about the
fundamental errors imposed by the noise but they have no information about the
corresponding error strengths. Hence, before Alice and Bob use the channel for
communication they would like to estimate the corresponding error rates. Then
they can, for example, decide on a suitable error correction scheme \cite{error}
or choose a
suitable encoding for their information \cite{signal}. In the remainder of the
paper we will
show that prior entanglement substantially increases the average reliability of
their estimation. \par 
\section{The Pauli Channel} 
The channel that we will investigate in this paper is the so called Pauli
channel $C$ which causes single qubit errors. 
These single qubit errors can be fully classified by the Pauli spin operators
$\hat{\sigma}_1=|0\rangle\langle 1|+|1\rangle\langle 0|$, 
$\hat{\sigma}_2=i(|1\rangle\langle 0|-|0\rangle\langle 1|)$ and
$\hat{\sigma}_3=|0\rangle\langle 0|-|1\rangle\langle 1|$ in the 
computational basis defined by $|0\rangle$ and $|1\rangle$.
The application of the unitary operators $\hat{\sigma}_i$ leads to a
fundamental rotation of a qubit state $|q\rangle=c_0 |0\rangle+c_1 |1\rangle$
with coefficients $c_i$. The \it bit flip \rm error is given by 
$\hat{\sigma}_1$, that is, $\hat{\sigma}_1 |q\rangle=c_0 |1\rangle+c_1
|0\rangle$. It exchanges the two basis states. The \it phase-flip \rm error
$\hat{\sigma}_3 |q\rangle=c_0 |0\rangle-c_1
|1\rangle$ changes the sign of $c_1$ in any coherent superposition of the basis states.
Finally $\hat{\sigma}_2$ generates a combination of bit and phase flip. \par
In a Pauli channel each of the three errors can occur with a certain probability
$p_i$ so that the superoperator reads
\begin{equation}
C(\hat{\rho})=\sum_{i=1}^4 p_i \hat{\sigma}_i \hat{\rho}\hat{\sigma}_i^{\dagger} 
\end{equation} 
with $\hat{\sigma}_4=\hat{1}$ and with probability $p_4=1-p_1-p_2-p_3$ that the
density operator remains unchanged. Hence the Pauli channel
is completely characterized by a parameter vector
$\vec{p}=(p_1,p_2,p_3)^T$.
Thus the action of the channel on the general
density operator 
\begin{equation}
\hat{\rho}(\vec{s})=\frac{1}{2}\left(\hat{1}+\sum_{i=1}^3 s_i\hat{\sigma}_i
\right)
\end{equation}
defined by the Bloch vector $\vec{s}=(s_1,s_2,s_3)^T$ with $s_i \in \Bbb{R}$ and $|\vec{s}|\le 1$
can be described by
the basic transformations
\begin{eqnarray}
C(\hat{\sigma}_1)&=&  \left[ 1-2(p_2+p_3)\right]\hat{\sigma}_1 \nonumber \\
C(\hat{\sigma}_2)&=&  \left[ 1-2(p_1+p_3)\right]\hat{\sigma}_2 \nonumber \\
C(\hat{\sigma}_3)&=&  \left[ 1-2(p_1+p_2)\right]\hat{\sigma}_3
\end{eqnarray} 
and $C(\hat{1})=\hat{1}$.
Our aim is to estimate the parameters $p_i$ from a finite amount of measurement
results. Hence the general scenario is the following: Alice prepares qubits in
well-known reference states and sends them to Bob through the channel to be
estimated. Bob knows those reference states and performs suited measurements on
the qubits he has received. The statistics of his measurement results will then
allow him to estimate the parameters $p_i$. \par
\section{Channel Estimation}
First, we consider the case -- also depicted in Fig. 1a -- that Alice sends 
single qubits through the channel.
In order to determine $\vec{p}$ Alice has to prepare three 
well-defined reference
states. For each of the states 
Bob then measures one operator so that at the end Bob has measured three
independent operators. \par
The natural choice is that Alice prepares three pure states (a)
$\hat{\rho}_1=\hat{\rho}(\vec{s}=(1,0,0)^T)$, 
(b) $\hat{\rho}_2=\hat{\rho}(\vec{s}=(0,1,0)^T)$, (c)
$\hat{\rho}_3=\hat{\rho}(\vec{s}=(0,0,1)^T)$ and Bob measures the operators (a)
$\hat{\sigma}_1$, (b) $\hat{\sigma}_2$ and (c) $\hat{\sigma}_3$
\cite{anmerkung}. The
corresponding expectation values $\langle\hat{\sigma}_i\rangle=1-2 P_i$ depend on the
probability $P_i$ to measure the eigenvalue $-1$ (spin down) in each case. With
the help of Eqs. (2) and (3) we immediately find that the parameter vector
\begin{equation}
\vec{p}=\frac{1}{2}\left(
\begin{array}{c}
P_3-P_1+P_2 \\
P_1-P_2+P_3 \\
P_2-P_3+P_1 
\end{array}
\right)
\end{equation}
can be calculated from the measured probabilities $P_i$. \par
If only finite resources are available Bob just finds frequencies instead of
probabilities. Using $M$ qubits for each of the three input states and the
corresponding measurements yields the
estimated parameters
 \begin{equation}
\vec{p}^{\; est}=\frac{1}{2M}\left(
\begin{array}{c}
i_3-i_1+i_2 \\
i_1-i_2+i_3 \\
i_2-i_3+i_1 
\end{array}
\right)
\end{equation}   
if $i_j$ results ``$-1$'' are recorded for the measurement of 
$\hat{\sigma}_j$. 
The reason why we choose the same number of $M$ qubits for each of these 
measurements is 
that we assume complete ignorance about the probabilities
$p_i$.\par
As the measure of estimation quality we use a standard statistical measure,
namely the quadratic deviation $\sum_{j=1}^3
(p_j-p_j^{est})^2$ which describes the error of the estimation. 
With this choice of cost function we find the average error
\begin{eqnarray}
\bar{f}(M,\vec{p})&=& \sum_{i_1=0}^M \sum_{i_2=0}^M \sum_{i_3=0}^M 
{M\choose i_1}{M\choose i_2}{M\choose i_3}(p_2+p_3)^{i_1} \nonumber \\
&\times &(1-p_2-p_3)^{M-i_1}
(p_1+p_3)^{i_2}(1-p_1-p_3)^{M-i_2}\nonumber \\
& \times & (p_1+p_2)^{i_3}(1-p_1-p_2)^{M-i_3}
\sum_{j=1}^3(p_j-p_j^{est})^2 \nonumber \\
&=& \frac{3}{2M} \left[p_1(1-p_1)+p_2(1-p_2)+p_3(1-p_3)\right. \nonumber \\
 &-& \left. p_1 p_2-p_2 p_3-p_1 p_3
\right]
\end{eqnarray}
averaged over all possible experimental outcomes. \par
Up to now we have only considered the case that Alice sends 
$N=3M$ single unentangled qubits through the
Pauli channel. However, one could also think of using entangled qubit pairs
(ebits) to
estimate the channel parameters. For this purpose we consider the following
second scenario which is also shown in Fig. 1b: 
Alice and Bob share an ebit prepared in a
$|\psi^-\rangle=1/\sqrt{2}(|0\rangle|1\rangle-|1\rangle|0\rangle)$ Bell state.
Alice sends her qubit through the Pauli channel whereas Bob simply keeps his
qubit. The channel causes the transformation
\begin{eqnarray}
C(|\psi^-\rangle\langle\psi^-|)&=& p_1|\phi^-\rangle\langle\phi^-|+
p_2|\phi^+\rangle\langle\phi^+| + p_3|\psi^+\rangle\langle\psi^+|\nonumber \\
&+& 
(1-p_1-p_2-p_3)|\psi^-\rangle\langle\psi^-|.
\end{eqnarray}
The Pauli channel transforms the initial $|\psi^-\rangle$ state into a
mixture of all four Bell states 
$|\phi^{\pm}\rangle=1/\sqrt{2}(|0\rangle|0\rangle\pm|1\rangle|1\rangle)$, 
$|\psi^{\pm}\rangle=1/\sqrt{2}(|0\rangle|1\rangle\pm|1\rangle|0\rangle)$ where
each Bell state is generated by exactly one of the possible single qubit errors
$\hat{\sigma}_{j}$.
Bob now performs a Bell measurement in which he finds each Bell state with
probability
\begin{eqnarray}
P_{|\phi^-\rangle}=&p_1& \qquad P_{|\phi^+\rangle}=p_2 
\qquad P_{|\psi^+\rangle}=p_3\nonumber \\
   & P_{|\psi^-\rangle}& =1-p_1-p_2-p_3.
\end{eqnarray}
Note that Alice and Bob can only use the same qubit resources as before. That
is, if they have used $N$ single qubits before they can now generate 
$N'=N/2$ ebits in the reference state $|\psi^-\rangle$. 
Consequently Bob gets only half as many measurement results as before: 
he finds four values $i_1$, $i_2$, $i_3$ and $i_4$
with $\sum_{j=1}^4 i_j=N'$ for the number of occurrences of
the Bell states $|\phi^-\rangle$, $|\phi^+\rangle$, $|\psi^+\rangle$ 
and $|\psi^-\rangle$. From these four values he
can calculate the estimated probabilities 
\begin{eqnarray}
p_1^{est}=\frac{i_1}{N'} &\qquad& p_2^{est}=\frac{i_2}{N'}\qquad
p_3^{est}=\frac{i_3}{N'}
\end{eqnarray}
that fully characterize the Pauli channel. 
The average error for the estimation scheme with entangled qubits then reads
\begin{eqnarray}
\bar{g}(N',\vec{p})&=&
\sum_{i_1+i_2+i_3+i_4=N'}\frac{N'!}{i_1!i_2!i_3!i_4!}\nonumber \\
&\times &
p_1^{i_1}p_2^{i_2}p_3^{i_3}(1-p_1-p_2-p_3)^{i_4}\sum_{j=1}^3(p_j-p_j^{est})^2 \nonumber \\
&=& \sum_{i_1=0}^{N'}\sum_{i_2=0}^{N'-i_1} \sum_{i_3=0}^{N'-i_1-i_2}
\frac{N'!}{i_1!i_2!i_3!(N'-i_1-i_2-i_3)!}\nonumber \\ 
&\times & p_1^{i_1}p_2^{i_2}p_3^{i_3}
(1-p_1-p_2-p_3)^{N'-i_1-i_2-i_3}
\nonumber \\ &\times & \sum_{j=1}^3(p_j-\frac{i_j}{N'})^2 
\nonumber \\ 
&=& \frac{1}{N'}\left[p_1(1-p_1)+p_2(1-p_2)+p_3(1-p_3)\right].
\end{eqnarray}
We can now compare the average errors, Eqs. (6) and (10), 
for both estimation schemes. As emphasized above for a fair comparison
we have to consider the same number $N=2N'=3M$ of available qubits 
for both schemes. 
We find that the difference 
\begin{eqnarray}
\Delta(N,\vec{p}) &=& \bar{f}(M=N/3,\vec{p})-\bar{g}(N'=N/2,\vec{p}) \nonumber \\
&=& \frac{1}{2N}\left[5(1-p_1-p_2-p_3)(p_1+p_2+p_3)\right. \nonumber \\
&+& \left.p_1 p_2+p_1 p_3+p_2 p_3\right] \ge 0
\end{eqnarray}
is non-negative for all possible parameter values $\vec{p}$.
This clearly shows that we indeed get an enhancement of the estimation quality due to
the use of entangled qubit pairs. This enhancement is illustrated in Fig. 2 for
the case $p_2=0$ which already shows the typical features of $\Delta$. We see
that $\Delta(N,\vec{p})$ is always positive except for the extremal points
$\vec{p}=(0,0,0)^T$, $\vec{p}=(1,0,0)^T$ and $\vec{p}=(0,0,1)^T$ where $\Delta$
vanishes.\par
Instead of comparing the errors for the same number of available qubits 
one could also compare the estimation quality for the same number $K$ of channel
applications. 
For the latter case we have $K=N'=3M$ so that the enhancement 
\begin{eqnarray}
\tilde{\Delta}(K,\vec{p}) 
&=& \bar{f}(M=K/3,\vec{p})-\bar{g}(N'=K,\vec{p}) \nonumber \\
&=& \frac{1}{2K}\left[7(1-p_1-p_2-p_3)(p_1+p_2+p_3)\right. \nonumber \\
&+& \left.5 p_1 p_2+5 p_1 p_3+5 p_2 p_3\right] \ge 0
\end{eqnarray}
due to
entanglement is even larger.\par

\section{Conclusion}
In conclusion we have shown a new application 
in which entanglement serves as a
superior resource for quantum information processing: In the case of the Pauli
channel there is a significant improvement 
in the estimation of the error strengths
due to the use of entangled qubits instead of separable ones. Hence
entanglement allows a 
better characterization of the quantum channel for finite initial resources
which can be counted in terms of available qubits or in terms of
channel applications. In contrast to dense coding \cite{denscode}, where we
achieve
an optimal encoding of information by entanglement, we obtain an enhanced
extraction of information about a quantum channel here. 
As
a consequence this additional information about the channel can be 
used in practical quantum communication problems to
optimize error correction schemes \cite{error} or signal ensembles 
\cite{signal}.  \par

\acknowledgments
We acknowledge discussions with G. Alber, A. Delgado, and M. Mussinger and 
support by the DFG programme
``Quanten-Informationsverarbeitung'',
by the European Science Foundation QIT programme and by the programmes
``QUBITS'' and ``QUEST'' of the European Commission.




\begin{figure}
\resizebox{8.6cm}{!}{\includegraphics{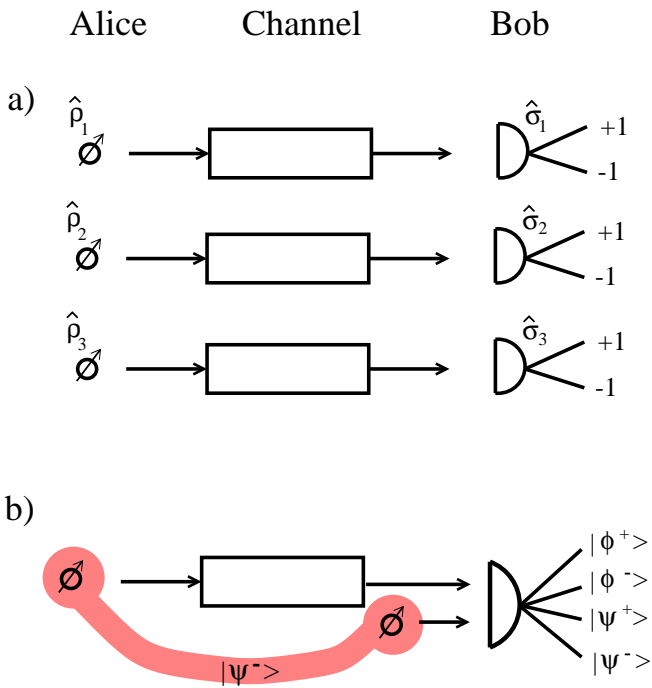}}
\vspace*{1cm}
\caption{The two schemes for estimating the
Pauli-channel parameters $p_i$ from an initial supply of $N$ qubits. 
In scheme a) Alice prepares single qubits in three different quantum states
$\hat{\rho}_i$ ($M=N/3$ qubits in each state) 
and sends them through the quantum channel. After receiving the
qubits Bob measures the operator $\hat{\sigma}_i$ for each qubit. Thus he
finally possesses $N$ measurement results from which he estimates the parameters
$p_i$. In scheme b) Alice and Bob share $N'=N/2$ entangled pairs of qubits
prepared in a $|\psi^-\rangle$ Bell state. Alice sends her qubit through the
quantum channel to Bob who then performs a Bell measurement onto the qubit
pairs.
In this scheme Bob records only $N/2$ measurement results.}
\end{figure} 

\begin{figure}
\resizebox{8.6cm}{!}{\includegraphics{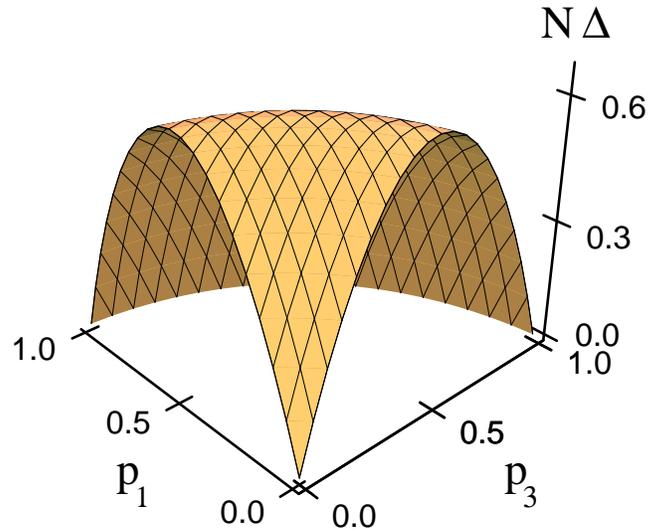}}
\caption{Plot of the estimation enhancement due to the use of entangled qubit
pairs. For the specific example of $p_2=0$ we have plotted $N\Delta$, Eq. (11), 
versus the
channel probabilities $p_1$ and $p_3$ in the allowed parameter range $0\le
p_1+p_3\le 1$. This quantity which is independent of $N$
is always non-negative. For $p_2=0$ the maximum gain of
$N\Delta_{max}=25/38\approx 0.66$ 
is reached for $p_1=p_3=5/19$. In the general case ($p_2\ge 0$) the maximum
value of $N\Delta_{max}=75/112\approx 0.67$ is found for $p_1=p_2=p_3=5/28$.}
\end{figure}


\begin{thebibliography}{99}
\bibitem{zeilinger} For a recent review see D. Bouwmeester, A. Ekert, and A.
Zeilinger (Eds.), \it The Physics of Quantum Information \rm (Springer, Berlin,
2000) and the references cited therein.
\bibitem{crypto} A.K. Ekert, Phys. Rev. Lett. \bf 67\rm, 661 (1991).
\bibitem{teleport} C.H. Bennett, G. Brassard, C. Crepeau, R. Jozsa, A. Peres,
and W.K. Wootters, Phys. Rev. Lett. \bf 70\rm, 1895 (1993); for more references
see Ref. \cite{zeilinger}.
\bibitem{denscode} C.H. Bennett and S.J. Wiesner, Phys. Rev. Lett. \bf 69\rm,
2881 (1992).
\bibitem{kraus} K. Kraus, \it States, Effects, and Operations\rm , Lecture Notes
in Physics Vol. 190 (Springer, Berlin, 1983).
\bibitem{schumacher} B. Schumacher, Phys. Rev. A \bf 51\rm, 2738 (1995).
\bibitem{shannon} G.E. Shannon and W. Weaver,\it The Mathematical Theory of
Communication \rm (University of Illinois Press, Urbana, 1949).
\bibitem{schuma1} B. Schumacher, Phys. Rev. A \bf 54\rm, 2614 (1996).
\bibitem{capacity} B.
Schumacher and M.A. Nielsen, Phys. Rev. A \bf 54\rm, 2629 (1996); C.H. Bennett,
D.P. DiVincenzo, J.A. Smolin, and W.K. Wootters, Phys. Rev. A \bf 54\rm, 3824
(1996); S. Lloyd, Phys. Rev. A \bf 55\rm, 1613 (1997); C.H. Bennett, D.P.
DiVincenzo, and J.A. Smolin, Phys. Rev. Lett. \bf 78\rm, 3217 (1997); B.
Schumacher and M.D. Westmoreland, Phys. Rev. A \bf 56\rm, 131 (1997); C. Adami
and N.J. Cerf, Phys. Rev. A \bf 56\rm, 3470 (1997); H. Barnum, M.A. Nielsen, and
B. Schumacher, Phys. Rev. A \bf 57\rm, 4153 (1998); D.P. DiVincenzo, P.W. Shor,
and J.A. Smolin, Phys. Rev. A \bf 57\rm, 830 (1998).
\bibitem{transclass} C.H. Bennett, P.W. Shor, J.A. Smolin, and A.V. Thapliyal,
Phys. Rev. Lett. \bf 83\rm, 1459 (1999).
\bibitem{analogue} H.K. Lo and S. Popescu, Phys. Rev. Lett. \bf 83\rm, 1459
(1999).
\bibitem{error} P.W. Shor, Phys. Rev. A \bf 52\rm, R2493 (1995); A.R. Calderbank
and P.W. Shor, Phys. Rev. A \bf 54\rm, 1098 (1996); R. Laflamme, 
C. Miquel, J.P. Paz, and W.H. Zurek, 
Phys. Rev. Lett. \bf 77\rm, 198 (1996); A.M. Steane, Phys. Rev. Lett.
\bf 77\rm, 793 (1996); E. Knill and R. Laflamme, Phys. Rev. A \bf 55\rm, 900
(1997).
\bibitem{signal} B. Schumacher and M.D. Westmoreland, e-print quant-ph/9912122
(1999).
\bibitem{anmerkung}	The
reason why this choice of prepared input states is optimal is that 
they are eigenstates of one
of the channel error operators $\hat{\sigma}_i$, cp. Eq. (3). Thus the
resulting output states only depend on two instead of three 
channel parameters $p_i$. The measurements are then designed in such a way that the
variation in measured probabilities $P_i$ due to a change of channel parameters 
is maximal.
 


\end{thebibliography}
\end{document}